\documentclass[runningheads]{llncs}
\usepackage{titletoc}
\usepackage{graphicx}
\usepackage{amsmath,amssymb,amsfonts}
\usepackage[misc]{ifsym}
\usepackage{subcaption}
\usepackage{csquotes}
\usepackage{textcomp}
\usepackage{float}
\usepackage{xcolor}
\usepackage[switch]{lineno}

\newcommand{\rl}{\mathbb{V}^L}

\newcommand{\inte}{\mathbb{Z}_L}
\newcommand{\complex}{\mathbb{C}^L}

\begin{document}
\title{Spark Deficient Gabor Frame Provides a Novel Analysis Operator for Compressed Sensing}
\titlerunning{SDGF provides a novel analysis operator for CS}

\author{Vasiliki Kouni\inst{1,2}\,{\Letter}\and
Holger Rauhut\inst{1}}
\authorrunning{V. Kouni and H. Rauhut}

\tocauthor{Vasiliki~Kouni, Holger~Rauhut}
\toctitle{Spark Deficient Gabor Frame Provides a Novel Analysis Operator for Compressed Sensing}

\institute{Chair for Mathematics of Information Processing, RWTH Aachen University, Germany\\ \email{$\{$kouni,rauhut$\}$@mathc.rwth-aachen.de} \and
Department of Informatics \& Telecommunications, National \& Kapodistrian University of Athens, Greece\\
\email{vicky-kouni@di.uoa.gr}\\}
\maketitle              
\begin{abstract}
The analysis sparsity model is a very effective approach in modern Compressed Sensing applications. Specifically, redundant analysis operators can lead to fewer measurements needed for reconstruction when employing the analysis $l_1$-minimization in Compressed Sensing. In this paper, we pick an eigenvector of the Zauner unitary matrix and --under certain assumptions on the ambient dimension-- we build a spark deficient Gabor frame. The analysis operator associated with such a frame, is a new (highly) redundant Gabor transform, which we use as a sparsifier in Compressed Sensing. We conduct computational experiments --on both synthetic and real-world data-- solving the analysis $l_1$-minimization problem of Compressed Sensing, with four different choices of analysis operators, including our Gabor analysis operator. The results show that our proposed redundant Gabor transform outperforms --in all cases-- Gabor transforms generated by state-of-the-art window vectors of time-frequency analysis.
\end{abstract}

\section{Introduction}
Compressed Sensing (CS) \cite{tao} is a modern technique to recover vectors $x\in\rl$ ($\mathbb{V}=\mathbb{R}$ or $\mathbb{C}$) from few linear and possibly corrupted measurements
\begin{equation}
    y=Ax+e\in\mathbb{V}^K,
\end{equation} $K<L$.
The applications of CS vary among Radar Imaging \cite{radar}, Cryptography \cite{crypto}, Telecommunications \cite{telecs}, Magnetic Resonance Imaging \cite{fistashear}, Deep Learning \cite{deepcs}.\\
\noindent\textbf{Related Work:} CS heavily relies on sparsity/compressibility of the signal of interest $x$.
Sparse data models are split in synthesis and analysis sparsity. 
The former is by now very well studied \cite{fistashear,rf,adcock,eeg}.
On the other hand, significant research has also been conducted over the last years towards its analysis counterpart \cite{genzel,kr,candes}, (also known as co-sparse model \cite{cosparse,kaba}), due to the flexibility it provides in  modelling sparse signals, since it leverages the redundancy of the involved analysis operators. Related work \cite{genzel} has also demonstrated that it is computationally more appealing to solve the optimization algorithm of analysis CS since a) the actual optimization takes place in the ambient space b) the algorithm may need less measurements for perfect reconstruction, if one uses a redundant transform instead of an orthogonal one.\\
\noindent\textbf{Motivation:} Our work is inspired by the articles \cite{genzel,kr,tvcs}, which propose either analysis operators associated to redundant frames
with atoms in general position, or a finite difference operator,
in which many linear dependencies appear for large dimensions. In a similar spirit, we also deploy frames, but we differentiate our approach by using \textit{spark deficient frames}, i.e. their elements are not in general linear position. Our intuition behind this choice is based on remarks of \cite{kaba}. The authors of \cite{kaba} refer to the union-of-subspaces model \cite{union}, according to which, it is desired to have analysis operators exhibiting high linear dependencies among their rows; this is a condition satisfied by spark deficient frames. To that end, we introduce a novel analysis operator associated with a spark deficient Gabor frame (SDGF). The latter can be generated by time-frequency shifts of any eigenvector of the Zauner unitary matrix \cite{zauner}, under certain assumptions. To the best of our knowledge, its efficiency when combined with CS has not yet been demonstrated. Moreover, since Gabor transforms are little explored in terms of CS \cite{eeg,tfh,gabmult}, we compare our proposed Gabor transform to three other Gabor transforms, emerging from state-of-the-art window vectors in time-frequency analysis. Finally, we illustrate the practical importance of our method for synthetic and real-world data.\\
\noindent\textbf{Key Contributions:} Our novelty is twofold: (a) we generate a SDGF based on a window vector, associate this SDGF to a new Gabor analysis operator and use the latter as a sparsifier in analysis CS (b) we compare numerically our proposed method with three other Gabor analysis operators, based on common windows of time-frequency analysis, on synthetic data and real-world speech signals. Our experiments show that our method outperforms all others, consistently for synthetic and real-world signals.


\section{Compressed Sensing setup}
\noindent\textbf{Notation:} For a set of indices $N=\{0,1,\dots,N-1\}$, we write $[N]$. The set of (column) vectors $|0\rangle,|1\rangle,\dots,|L-1\rangle$ is the standard basis of $\mathbb{C}^L$. We write $\inte$ for the ring of residues $\mathrm{mod}L$, that is $\inte=\{0\mathrm{mod}L,1\mathrm{mod}L,\dots,(L-1)\mathrm{mod}L\}$ and $a\equiv b(\mathrm{mod}L)$ denotes the congruence modulo, $a,\,b\in\mathbb{Z}$. The support of a signal $x\in\rl$ is denoted by $\mathrm{supp}(x)=\{i\in[L]:x_i\neq0\}$. For its cardinality we write $|\mathrm{supp}(x)|$ and if $|\mathrm{supp}(x)|\leq s<<L$, we call $x$ $s$-sparse.

\noindent\textbf{Analysis Compressed Sensing Formulation:} As already described in Section I, the main idea of CS is to reconstruct a signal $x\in\mathbb{V}^L$ from $y=Ax+e\in\mathbb{V}^K$, $K<L$, where $A$ is the so-called measurement matrix and $e\in\mathbb{V}^K$, with $\|e\|_2\leq\eta$, corresponds to noise. To do so, we first assume there exists a redundant sparsifying transform $\Phi\in\mathbb{V}^{P\times L}$ ($P>L$) called the analysis operator, such that $\Phi x$ is (approximately) sparse. On the other hand, the choice of $A$ is tailored to the application for which CS is employed. In this paper, we choose $A$ to be a randomly subsampled identity operator, since this is considered a standard CS setup. Moreover, this type of measurement matrix has proven to work well \cite{adcock}, since it meets conditions ensuring exact or approximate reconstruction of $x$, i.e., the matrix has small \textit{coherence} or satisfies the \textit{restricted isometry property} \cite{rf}.
Now, using analysis sparsity in CS, we wish to recover $x$ from $y$. A common approach is the \textit{analysis $l_1$-minimization} problem
\begin{equation}
    \label{denl1}
        \min_{x\in\rl}\|\Phi x\|_1\quad\text{subject to}\quad \|Ax-y\|_2\leq \eta,
\end{equation}
or a regularized\footnote{in terms of optimization, it is preferred to solve \eqref{denl1} instead of \eqref{regl1}} version \cite{tfocs} of it:
\begin{equation}
\label{regl1}
        \min_{x\in\rl}\|\Phi x\|_1+\frac{\mu}{2}\|x-x_0\|_2^2\quad\text{subject to}\quad \|Ax-y\|_2\leq \eta,
\end{equation} with $x_0$ being an initial guess on $x$ and $\mu>0$ a smoothing parameter. We will devote the next Section to the construction of a suitable analysis operator $\Phi$.

\section{Gabor Frames}
\noindent\textbf{Gabor Systems:} A \textit{discrete Gabor system} $(g,a,b)$ \cite{dgs} is defined as a collection of time-frequency shifts of the so-called window vector $g\in\mathbb{C}^L$, expressed as
\begin{align}
    \label{gaborsystem}
    g_{n,m}(l)&=e^{2\pi imbl/L}g(l-na),\quad l\in[L],
\end{align}
where $a,\,b$ denote time and frequency (lattice) parameters respectively, $n\in[N]$ chosen such that $N=L/a\in\mathbb{N}$ and $m\in[M]$ chosen such that $M=L/b\in\mathbb{N}$ denote time and frequency shift indices, respectively. If \eqref{gaborsystem} spans $\mathbb{C}^L$, it is called a \textit{Gabor frame}.
The number of elements in $(g,a,b)$ according to \eqref{gaborsystem} is $P=MN=L^2/ab$ and if $(g,a,b)$ is a frame, we have $ab<L$. Good time-frequency resolution of a signal with respect to a Gabor frame, depends on appropriately choosing $a$, $b$. This challenge can only be treated by numerically experimenting with different values of $a,\,b$ with respect to $L$. Now, we associate to the Gabor frame $(g,a,b)$ the following operator.
\begin{definition}
Let $\Phi_g:\complex\mapsto\mathbb{C}^{M\times N}$ denote the \textit{Gabor analysis operator} --also known as \textit{digital Gabor transform} (DGT)-- whose action on a signal $x\in\complex$ is defined as
\begin{equation}\label{gabcoeff}
    c_{m,n}=\sum_{l=0}^{L-1}x_l\overline{g(l-na)}e^{-2\pi imbl/L},\qquad m\in[M],\,n\in[N].
\end{equation}

\end{definition}
\noindent\textbf{Spark Deficient Gabor Frames:} Let us first introduce some basic notions.
\begin{definition}
The symplectic group $\mathrm{SL}(2,\inte)$ consists of all matrices
\begin{equation}
    G=\begin{pmatrix}
\alpha & \beta\\
\gamma & \delta
\end{pmatrix}
\end{equation}
such that $\alpha,\,\beta,\,\gamma,\,\delta\in\inte$ and $\alpha\delta-\beta\gamma\equiv1(\mathrm{mod}L)$. To each such matrix  corresponds a unitary matrix given by the explicit formula \cite{dang}
\begin{equation}\label{sic}
    U_G=\frac{e^{i\theta}}{\sqrt{L}}\sum_{u,v=0}^{L-1}\tau^{\beta^{-1}(\alpha v^2-2uv+\delta u^2)}|u\rangle\langle v|,
\end{equation}
where $\theta$ is an arbitrary phase, $\beta^{-1}$ is the inverse\footnote{$bb^{-1}\equiv1(\mathrm{mod}L)$} of $\beta\mathrm{mod}L$ and $\tau=-e^{\frac{i\pi}{L}}$.
\end{definition}
\begin{definition}
The \textit{spark} of a set $F$ --denoted by $\mathrm{sp}(F)$-- of $P$ vectors in $\complex$ is the size of the smallest linearly dependent subset of $F$. A frame $F$ is full spark if and only if every set of $L$ elements of $F$ is a basis, or equivalently $\mathrm{sp}(F)=L+1$, otherwise it is spark deficient.
\end{definition}

Based on the previous definition, a Gabor frame with $P=L^2/ab$ elements of the form \eqref{gaborsystem} is full spark, if and only if every set of $L$ of its elements is a basis. Now, as proven in \cite{mal}, almost all window vectors generate full spark Gabor frames, so the SDGFs are generated by exceptional window vectors.  Indeed, the following theorem was proven in \cite{dang} and informally stated in \cite{sparkmal}, for the \textit{Zauner} matrix $\mathcal{Z}\in\mathrm{SL}(2,\inte)$ given by
\begin{equation}\label{zauner}
    \mathcal{Z}=\begin{pmatrix}
0 & -1\\
1 & -1
\end{pmatrix}\equiv\begin{pmatrix}
0 & L-1\\
1 & L-1
\end{pmatrix}.
\end{equation}
\begin{theorem}[\cite{dang}]\label{sdgf}
Let $L\in\mathbb{Z}$ such that $2\nmid L$, $3\mathrel{|} L$ and $L$ is square-free. Then, any eigenvector of the Zauner unitary matrix $U_\mathcal{Z}$ (produced by combining \eqref{sic} and \eqref{zauner}), generates a spark deficient Gabor frames for $\complex$.
\end{theorem}
According to Theorem \ref{sdgf}, since all the eigenvectors of $U_\mathcal{Z}$ generate SDGFs, we may choose without loss of generality an arbitrary one, call it \textit{star window vector} and denote it as $g_*$. We call \textit{star-DGT} the analysis operator associated with a SDGF produced by $g_*$, and denote it $\Phi_{g_*}$. We coin the term "star", due to the slight resemblance of this DGT to a star when plotted in MATLAB.
\begin{remark}\label{lchoice}
A simple way to choose $L$, is by considering its prime factorization: take $k$ prime numbers $p_1^{\alpha_1},\dots,p_k^{\alpha_k}$, with $\alpha_1,\dots,\alpha_k$ not all a multiple of 2 and $p_1=3,p_i\neq2,i=2,\dots,k$, such that $L=3^{\alpha_1}p_2^{\alpha_2}\cdot\dots\cdot p_k^{\alpha_k}$. Since $a,b\mathrel{|}L$, we may also choose $a=1$ and $b=p_i^{\alpha_i},i=1,\dots,k$. Otherwise, both $a,b$ may be one, or a multiplication of more than one, prime numbers from the prime factorization of $L$. We have seen empirically that this method for fixing $(L,a,b)$ produces satisfying results, as it is illustrated in the figures of the upcoming pages.
\end{remark}


\section{Numerical Experiments}
\noindent\textbf{Signals' description and preprocessing:} We experiment with 3 synthetic data and 6 real-world speech signals, taken from Wavelab package \cite{wavelab} and TIMIT corpus \cite{timit}, respectively. All signals are real-valued; the real-world data are sampled at 16 kHz. The true ambient dimension of each real-world signal does not usually match the conditions of Theorem \ref{sdgf}. Hence, we use Remark \ref{lchoice} to cut-off each speech signal to a specific ambient dimension $L$, being as closer as it gets to its true dimension, in order to both capture a meaningful part of the signal and meet the conditions of Theorem \ref{sdgf}. For the synthetic data, we use again Theorem \ref{sdgf} and Remark \ref{lchoice} to fix each signal's ambient dimension $L$.

\begin{table}[htbp]
    \caption{Signals' details and summary of parameters}
    \centering
    \scalebox{0.8}{\begin{tabular}{|| c | c | c | c | c ||}
    \hline
     Labels & Samples & $(L,a,b)$ & $x_0$ & $\mu_i$, $i=1,2,3,*$\\
     \hline\hline
     Cusp & 33 & $(33,1,11)$ & zero vector & $\|\Phi_ix\|_\infty$\\
     \hline
     Ramp & 33 & $(33,1,11)$ & zero vector & $\|\Phi_ix\|_\infty$\\
     \hline
     Sing & 45 & $(45,1,9)$ & zero vector & $\|\Phi_ix\|_\infty$\\
     \hline
     SI1899 & 22938 & $(20349,19,21)$ & $A^TAx$ & $10^{-1}\|\Phi_ix\|_\infty$\\
     \hline
     SI1948 & 27680 & $(27531,19,23)$ & $A^TAx$ & $10^{-1}\|\Phi_ix\|_\infty$\\
     \hline
     SI2141 & 42800 & $(41769,21,17)$ & $A^TAx$ & $10^{-1}\|\Phi_ix\|_\infty$\\
     \hline
     SX5 & 24167 & $(23205,17,13)$ & $A^TAx$ & $10^{-1}\|\Phi_ix\|_\infty$\\
     \hline
     SX224 & 25805 & $(24633,23,21)$ & $A^TAx$ & $10^{-1}\|\Phi_ix\|_\infty$\\
     \hline
     SI1716 & 25908 & $(24633,23,21)$ & $A^TAx$ & $\|\Phi_ix\|_\infty$\\
     \hline
    \end{tabular}}
    \label{sigdec}
\end{table}

\noindent\textbf{Proposed framework for each signal:}
We choose $a,b$ according to Remark \ref{lchoice}. We consider a vector $K$ of 1000 evenly spaced points in $[1,L]$ and use it as the measurements' interval. We use the power iteration method \cite{power} which yields the largest in magnitude eigenvalue and  corresponding eigenvector of  $U_{\mathcal{Z}}$, then set this eigenvector to be the star window vector. We use the MATLAB package \textit{LTFAT} \cite{ltfat}, to generate four different Gabor frames, with their associated analysis operators/DGTs: $\Phi_{g_1}$, $\Phi_{g_2}$, $\Phi_{g_3}$ and  $\Phi_{g_*}$, corresponding to a Gaussian, a Hann, a Hamming \cite{dgs} and the star window vector, respectively. Since we process real-valued signals, we alter the four analysis operators to compute only the DGT coefficients of positive frequencies. For each choice of $K$ in the measurements' interval, we set up a randomly subsampled identity operator $A\in\mathbb{R}^{K\times L}$ and determine the noisy measurements $y=Ax+e$, with $e$ being zero-mean Gaussian noise with standard deviation $\sigma=0.001$. We employ the MATLAB package \textit{TFOCS} \cite{tfocs} to solve four different instances of \eqref{regl1}, one for each of the four DGTs. For each $\Phi_{g_i}$, $i=1,2,3,*$, we set $\mu_i=C\|\Phi_ix\|_\infty$, $C>0$, since we noticed an improved performance of the algorithm when $\mu$ is a function of $\Phi_i$ (the constant $C$ and the function $\|\cdot\|_\infty$ are simply chosen empirically). From the aforementioned procedure, we obtain four different estimators of $x$, namely $\hat{x}_1$, $\hat{x}_2$, $\hat{x}_3$,  $\hat{x}_*$ and their corresponding relative errors $\|x-\hat{x}_i\|_2/\|x\|_2$, $i=1,2,3,*$.

\begin{figure*}[h!]
\centering
\begin{subfigure}
{.3\linewidth}
\centering
\includegraphics[width=\textwidth]{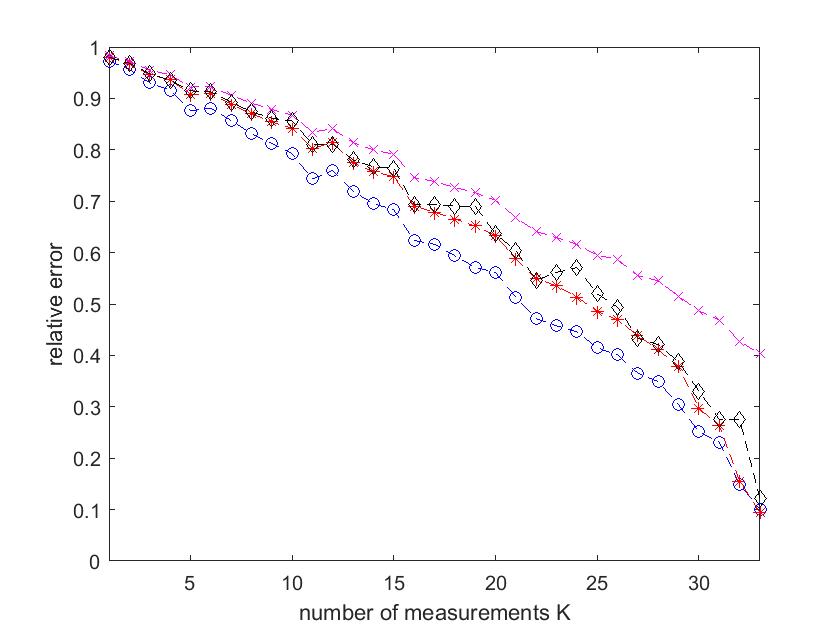}
\captionsetup{justification=centering}
\caption{Cusp with $(L,a,b)=(33,1,11)$}
\label{cusp}
\end{subfigure}\hfill
\begin{subfigure}
{.3\linewidth}
\centering
\includegraphics[width=\textwidth]{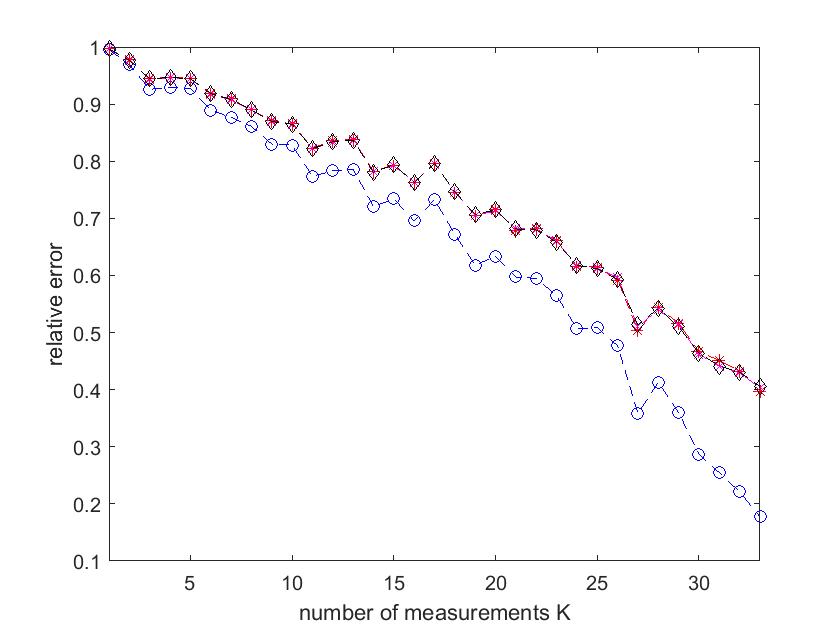}
\captionsetup{justification=centering}
\caption{Ramp with $(L,a,b)=(33,1,11)$}
\label{ramp}
\end{subfigure}\hfill
\begin{subfigure}
{.3\linewidth}
\centering
\includegraphics[width=\textwidth]{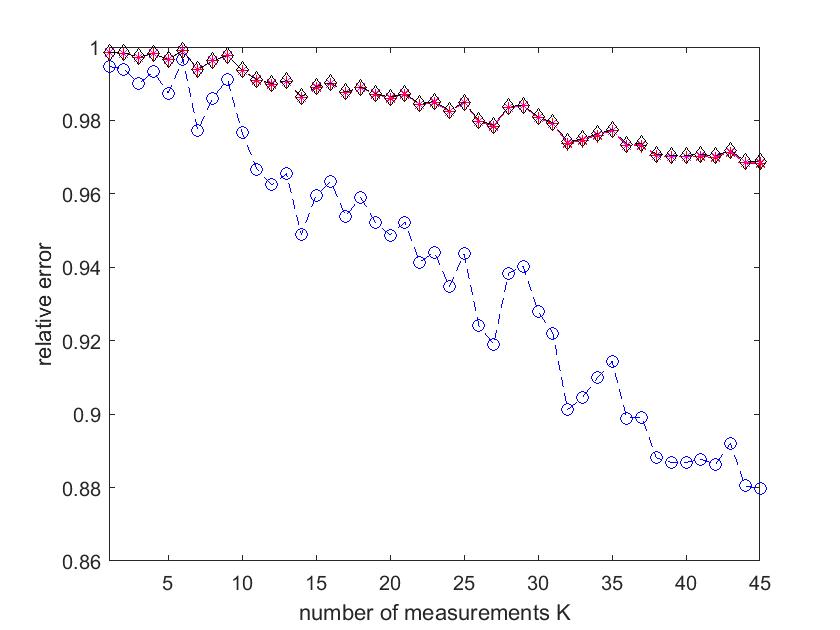}
\captionsetup{justification=centering}
\caption{Sing with $(L,a,b)=(45,1,9)$}
\label{sing}
\end{subfigure}\vfill
\begin{subfigure}
{.5\linewidth}
\centering
\includegraphics[width=\linewidth]{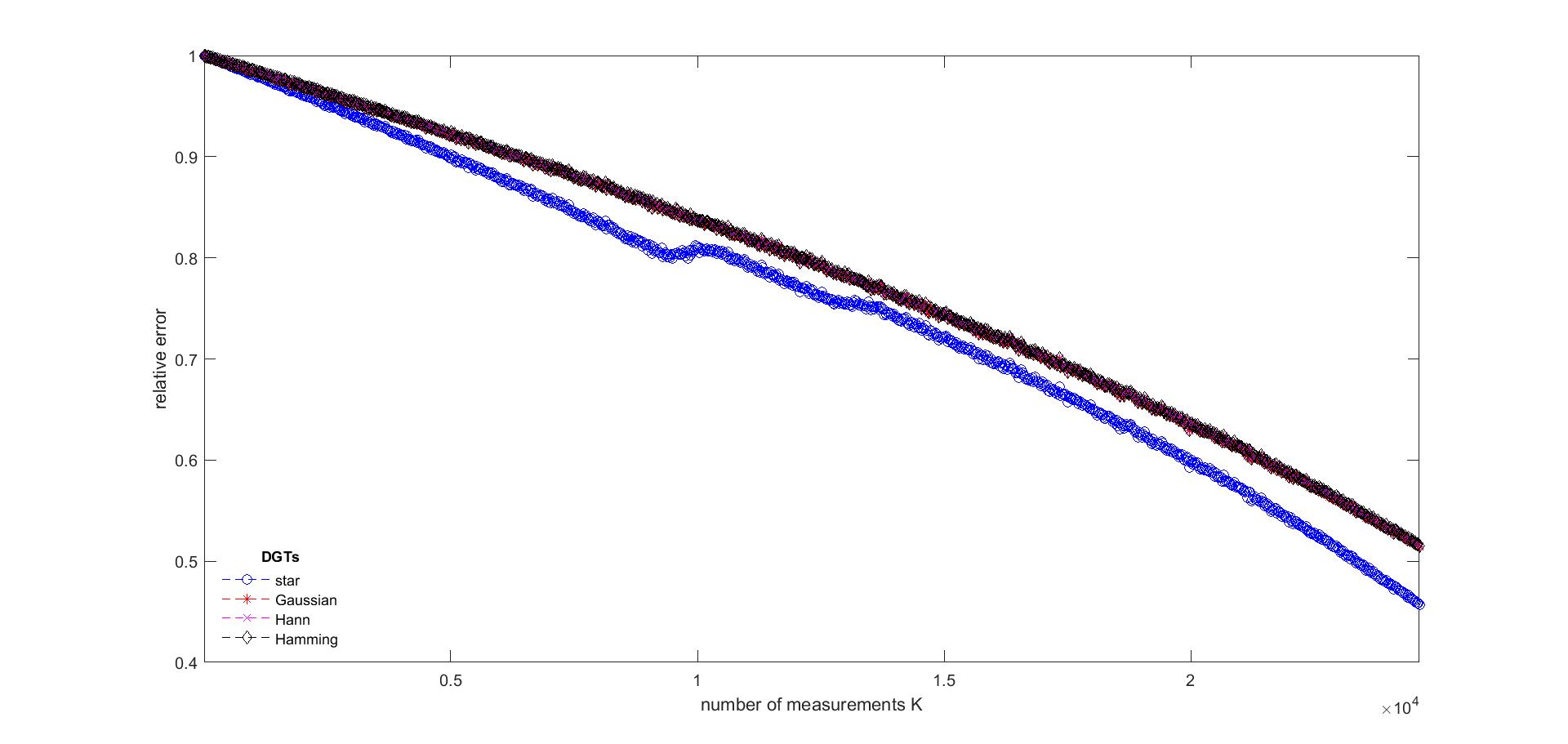}
\captionsetup{justification=centering}
\caption{SI1716 with $(L,a,b)=(24633,23,21)$}
\label{si1716}
\end{subfigure}\hfill
\begin{subfigure}
{.5\linewidth}
\centering
\includegraphics[width=\linewidth]{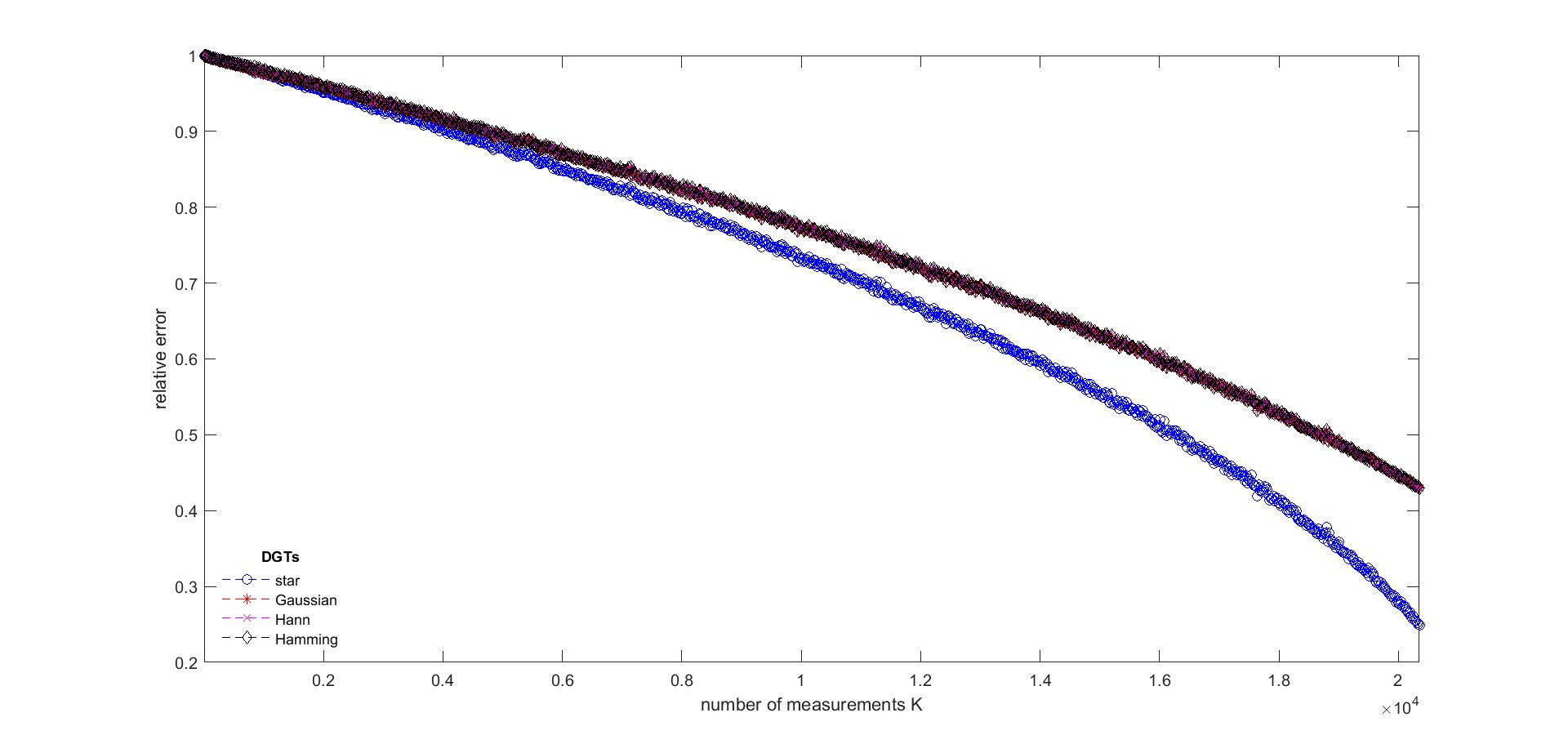}
\captionsetup{justification=centering}
\caption{SI1899 with $(L,a,b)=(20349,19,21)$}
\label{si1899}
\end{subfigure}\vfill
\begin{subfigure}
{.5\linewidth}
\centering
\includegraphics[width=\textwidth]{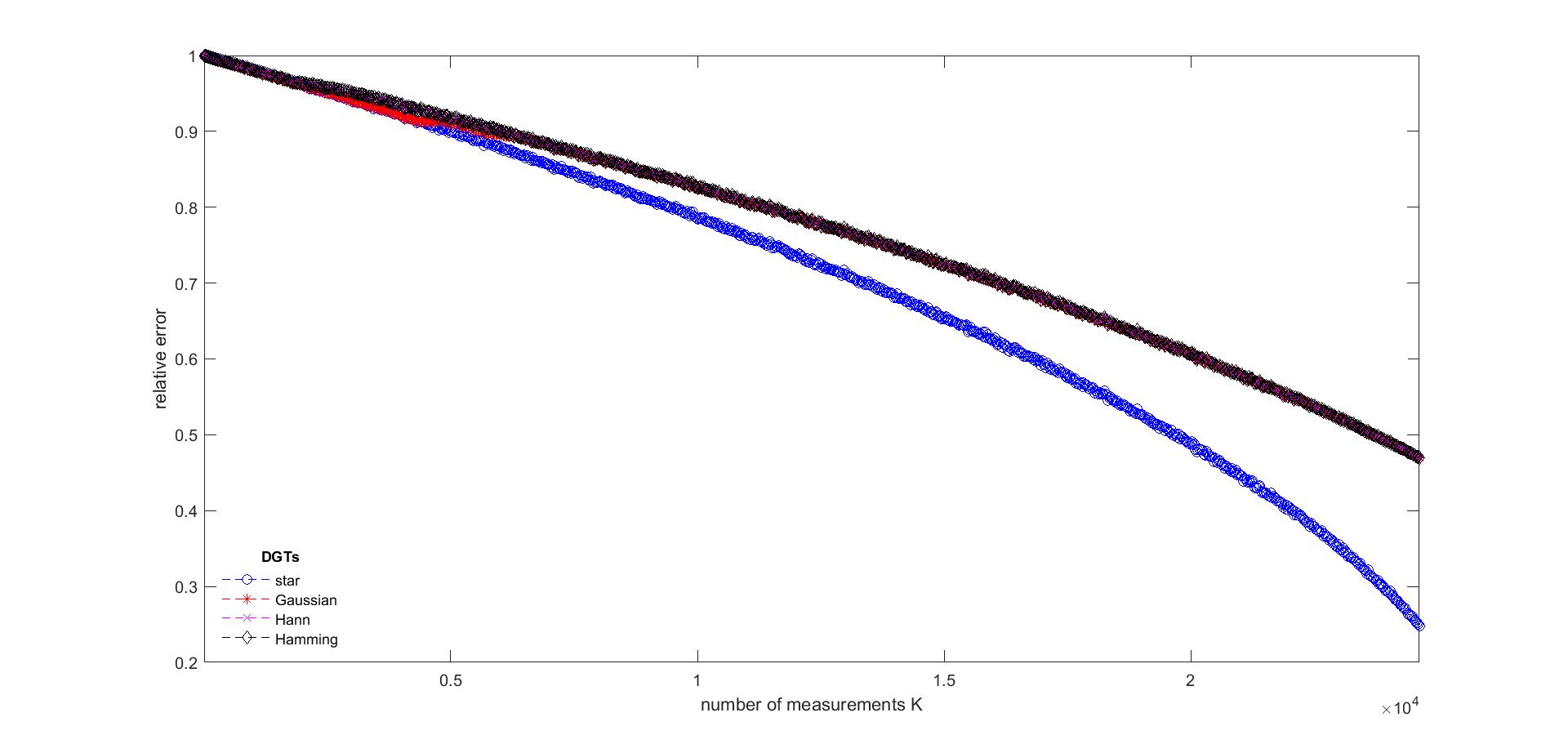}
\captionsetup{justification=centering}
\caption{SI1948 with $(L,a,b)=(24633,21,23)$}
\label{si1948}
\end{subfigure}\hfill
\begin{subfigure}
{.5\linewidth}
\centering
\includegraphics[width=\textwidth]{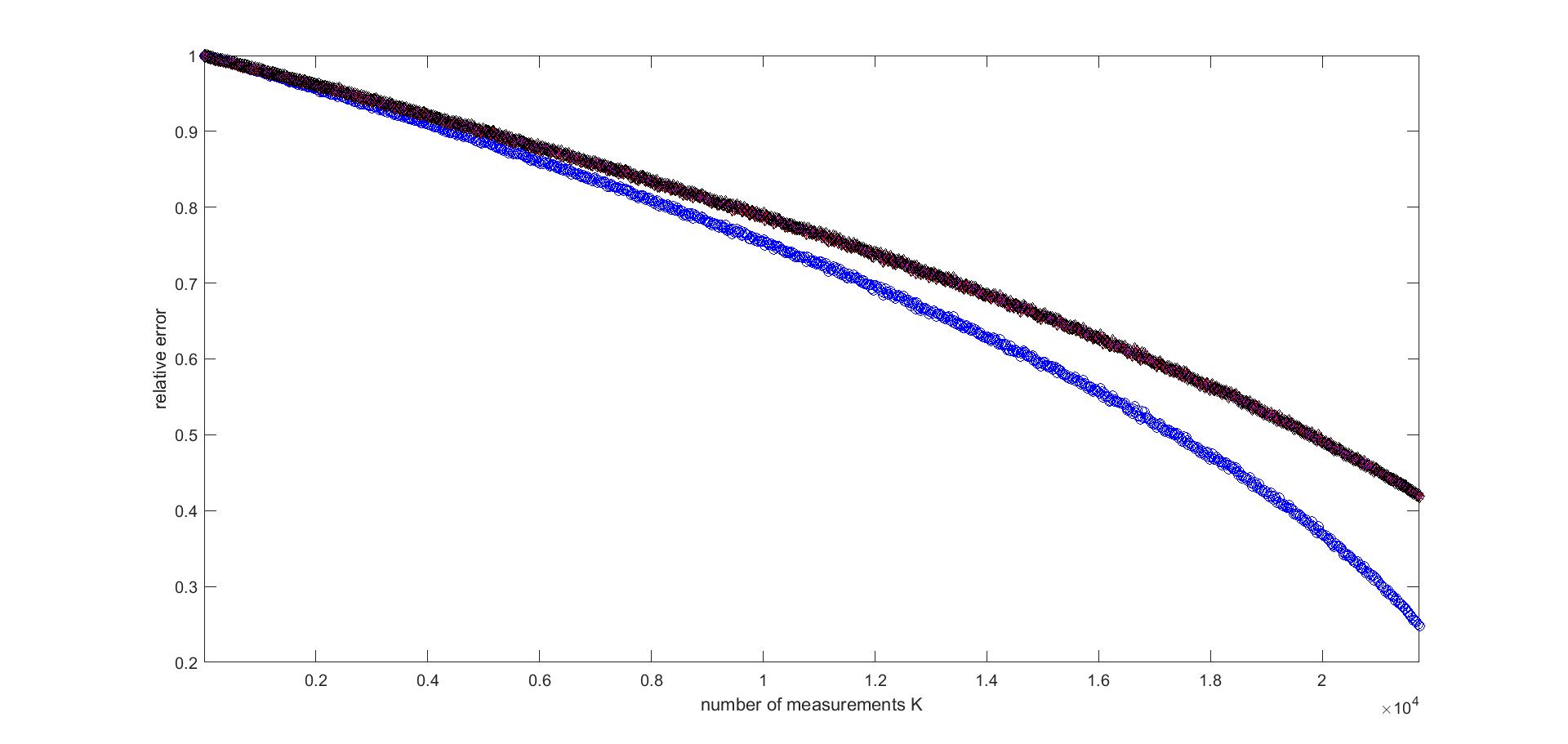}
\captionsetup{justification=centering}
\caption{SI2141 with $(L,a,b)=(21735,21,23)$}
\label{si2141}
\end{subfigure}\vfill
\begin{subfigure}
{.5\linewidth}
\centering
\includegraphics[width=\textwidth]{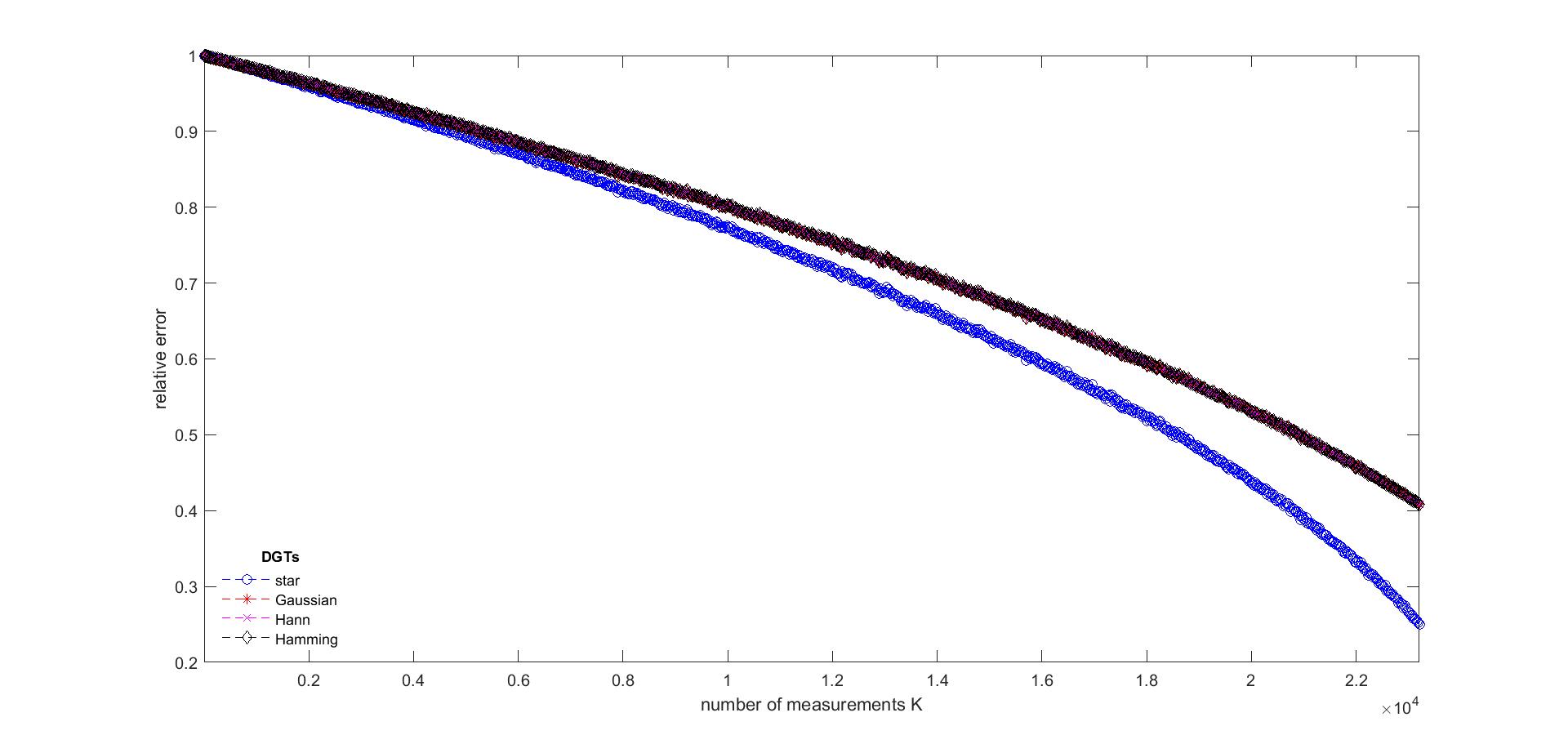}
\captionsetup{justification=centering}
\caption{SX5 with $(L,a,b)=(23205,17,13)$}
\label{sx5}
\end{subfigure}\hfill
\begin{subfigure}
{.5\linewidth}
\centering
\includegraphics[width=\textwidth]{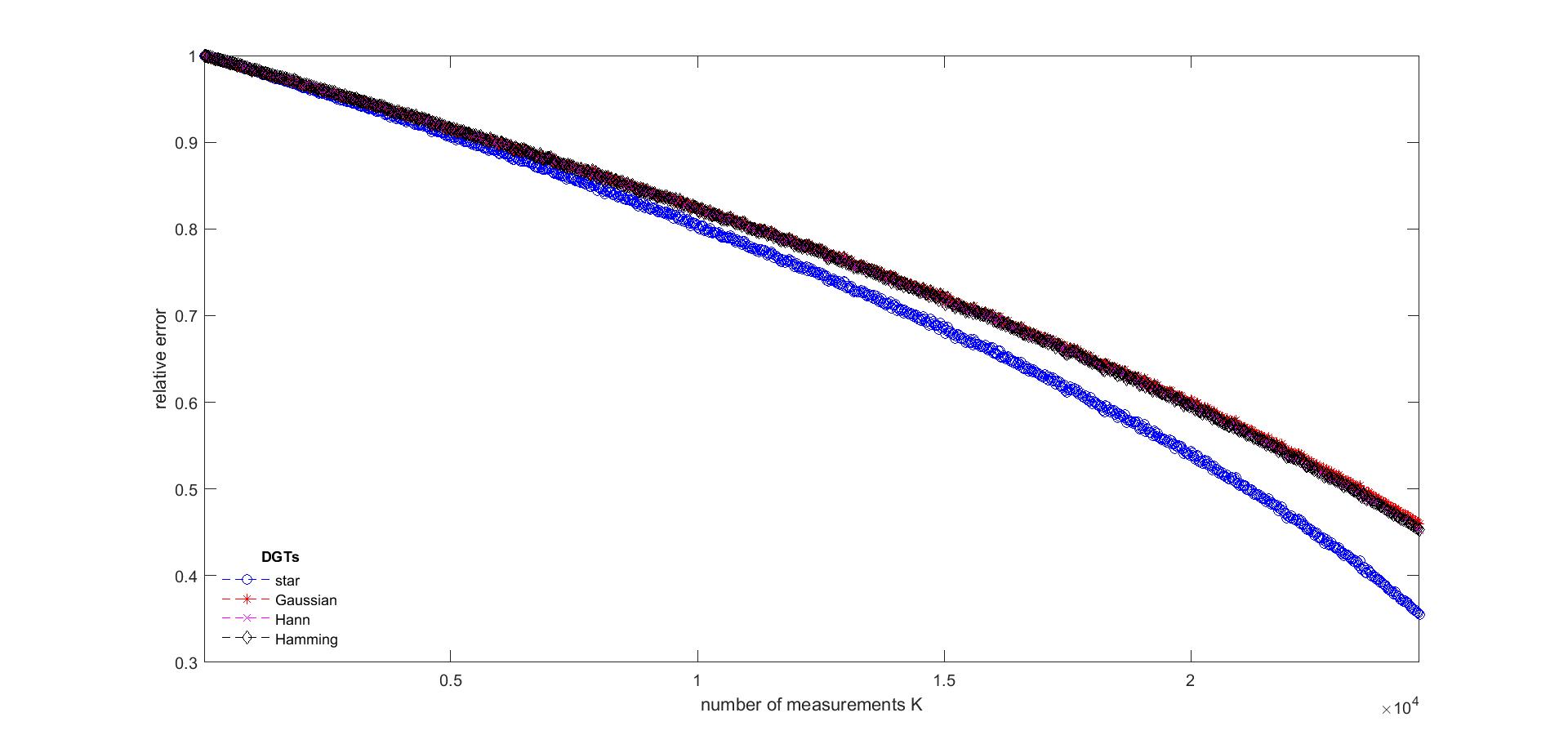}
\captionsetup{justification=centering}
\caption{SX224 with $(L,a,b)=(24633,23,21)$}
\label{sx224}
\end{subfigure}
\captionsetup{justification=centering}
\caption{Rate of approximate success for synthetic ((a)-(c)) and real-world ((d)-(i)) signals for different parameters $(L,a,b)$. Red: Gaussian, magenta: Hann, black: Hamming, blue: proposed.}
\label{realworld}
\end{figure*}

\noindent\textbf{Discussion of the results:} The labels of all signals, along with short description and some key characteristics of the application of our framework to all signals, can be found in Table~\ref{sigdec}. The resulting figures show the relative reconstruction error decay as the number of measurements increases. Fig.~\ref{cusp}--\ref{sing} demonstrate the success rate of our proposed DGT (blue line), outperforming the rest of DGTs for the synthetic data. Similarly,  for the real-world speech signals, Fig.~\ref{si1716}--\ref{sx224} indicate that our method (again blue line) achieves state-of-the-art performance in the first $15-20\%$ of the measurements and from this point on, star-DGT outperforms the rest of DGTs. Moreover, the TFOCS algorithm needed only one iteration to reconstruct the signal of interest when our proposed star-DGT was employed; for the rest 3 DGTs, the algorithm needed at least three iterations. This behaviour confirms improved performance when a DGT associated with a SDGF is applied to analysis CS.

\section{Conclusion and Future Work}
In the present paper, we took advantage of a window vector to generate a spark deficient Gabor frame and introduced a novel redundant analysis operator/DGT, namely the star DGT, associated with this SDGF. We then applied the star DGT to analysis Compressed Sensing, along with three other DGTs generated by state-of-the-art window vectors in the field of Gabor Analysis. Our experiments confirm improved performance: the increased amount of linear dependencies provided by this SDGF, yields in all cases lower relative reconstruction error for both synthetic and real-world data, as the number of measurements increases. Future directions will be the extension of the present framework to largescale problems (e.g. images or videos). Additionally, it would be interesting to compare this star-DGT to other similar choices of redundant analysis operators (e.g. redundant wavelet transform, shearlets \cite{shear} etc.).
%
%
%


\begin{thebibliography}{8}
\bibitem{tao}
Candès, E.J., Romberg, J. and Tao, T., 2006. Robust uncertainty principles: Exact signal reconstruction from highly incomplete frequency information. IEEE Transactions on information theory, 52(2), pp. 489-509.

\bibitem{radar}
Potter, L.C., Ertin, E., Parker, J.T. and Cetin, M., 2010. Sparsity and compressed sensing in radar imaging. Proceedings of the IEEE, 98(6), pp. 1006-1020.

\bibitem{crypto}
Chen, J., Zhang, Y., Qi, L., Fu, C. and Xu, L., 2018. Exploiting chaos-based compressed sensing and cryptographic algorithm for image encryption and compression. Optics and Laser Technology, 99, pp. 238-248.

\bibitem{telecs}
Alexandropoulos, G.C. and Chouvardas, S., 2016, December. Low complexity channel estimation for millimeter wave systems with hybrid A/D antenna processing. In 2016 IEEE Globecom Workshops (GC Wkshps) (pp. 1-6). IEEE.

\bibitem{fistashear}
Pejoski, S., Kafedziski, V. and Gleich, D., 2015. Compressed sensing MRI using discrete nonseparable shearlet transform and {FISTA}. IEEE Signal Processing Letters, 22(10), pp.1566-1570.

\bibitem{deepcs}
Wu, Y., Rosca, M. and Lillicrap, T., 2019, May. Deep compressed sensing. In International Conference on Machine Learning (pp. 6850-6860). {PMLR}.

\bibitem{rf}
Foucart, S. and Rauhut, H., 2013. An invitation to compressive sensing. In A mathematical introduction to compressive sensing (pp. 1-39). Birkhäuser, New York, NY.

\bibitem{adcock}
Li, C. and Adcock, B., 2019. Compressed sensing with local structure: uniform recovery guarantees for the sparsity in levels class. Applied and Computational Harmonic Analysis, 46(3), pp.453-477.

\bibitem{eeg}
Dao, P.T., Griffin, A. and Li, X.J., 2018, July. Compressed sensing of EEG with Gabor dictionary: Effect of time and frequency resolution. In 2018 40th annual international conference of the ieee engineering in medicine and biology society (EMBC) (pp. 3108-3111). IEEE.

\bibitem{genzel}
Genzel, M., Kutyniok, G. and März, M., 2021. $l_1$-Analysis minimization and generalized (co-) sparsity: When does recovery succeed?. Applied and Computational Harmonic Analysis, 52, pp. 82-140.

\bibitem{kr}
Kabanava, M. and Rauhut, H., 2015. Analysis $l_1$-recovery with frames and gaussian measurements. Acta Applicandae Mathematicae, 140(1), pp. 173-195.

\bibitem{candes}
Candes, E.J., Eldar, Y.C., Needell, D. and Randall, P., 2011. Compressed sensing with coherent and redundant dictionaries. Applied and Computational Harmonic Analysis, 31(1), pp. 59-73.

\bibitem{cosparse}
Nam, S., Davies, M.E., Elad, M. and Gribonval, R., 2013. The cosparse analysis model and algorithms. Applied and Computational Harmonic Analysis, 34(1), pp. 30-56.

\bibitem{kaba}
Kabanava, M. and Rauhut, H., 2015. Cosparsity in compressed sensing. In Compressed Sensing and Its Applications (pp. 315-339). Birkhäuser, Cham.

\bibitem{tvcs}
Krahmer, F., Kruschel, C. and Sandbichler, M., 2017. Total variation minimization in compressed sensing. In Compressed Sensing and its Applications (pp. 333-358). Birkhäuser, Cham.

\bibitem{union}
Blumensath, T. and Davies, M.E., 2009. Sampling theorems for signals from the union of finite-dimensional linear subspaces. IEEE Transactions on Information Theory, 55(4), pp. 1872-1882.

\bibitem{zauner}
Zauner, G., 1999. Quantum designs (Doctoral dissertation, University of Vienna, Vienna).

\bibitem{tfh}
Pfander, G.E. and Rauhut, H., 2010. Sparsity in time-frequency representations. Journal of Fourier Analysis and Applications, 16(2), pp. 233-260.

\bibitem{gabmult}
Rajbamshi, S., Tauböck, G., Balazs, P. and Abreu, L.D., 2019, September. Random Gabor multipliers for compressive sensing: a simulation study. In 2019 27th European Signal Processing Conference (EUSIPCO) (pp. 1-5). IEEE.


\bibitem{tfocs}
Becker, S.R., Candès, E.J. and Grant, M.C., 2011. Templates for convex cone problems with applications to sparse signal recovery. Mathematical programming computation, 3(3), p. 165.

\bibitem{dgs}
Søndergaard, P.L., Hansen, P.C. and Christensen, O., 2007. Finite discrete Gabor analysis (Doctoral dissertation, Institut for Matematik, DTU).

\bibitem{dang}
Dang, H.B., Blanchfield, K., Bengtsson, I. and Appleby, D.M., 2013. Linear dependencies in Weyl–Heisenberg orbits. Quantum information processing, 12(11), pp. 3449-3475.

\bibitem{mal}
Malikiosis, R.D., 2015. A note on Gabor frames in finite dimensions. Applied and Computational Harmonic Analysis, 38(2), pp. 318-330.

\bibitem{sparkmal}
Malikiosis, R.D., 2018. Spark deficient Gabor frames. Pacific Journal of Mathematics, 294(1), pp. 159-180.

\bibitem{wavelab}
Buckheit, J.B. and Donoho, D.L., 1995. Wavelab and reproducible research. In Wavelets and statistics (pp. 55-81). Springer, New York, NY.

\bibitem{timit}
Garofolo, J.S., Lamel, L.F., Fisher, W.M., Fiscus, J.G. and Pallett, D.S., 1993. DARPA TIMIT acoustic-phonetic continous speech corpus CD-ROM. NIST speech disc 1-1.1. NASA STI/Recon technical report n, 93, p. 27403.

\bibitem{power}
Booth, T.E., 2006. Power iteration method for the several largest eigenvalues and eigenfunctions. Nuclear science and engineering, 154(1), pp. 48-62.

\bibitem{ltfat}
Pruša, Z., Søndergaard, P., Balazs, P. and Holighaus, N., 2013, October. LTFAT: A Matlab/Octave toolbox for sound processing. In Proc. 10th International Symposium on Computer Music Multidisciplinary Research (CMMR) (pp. 299-314).

\bibitem{shear}
Yuan, M., Yang, B., Ma, Y., Zhang, J., Zhang, R. and Zhang, C., 2015. Compressed sensing MRI reconstruction from highly undersampled-space data using nonsubsampled shearlet transform sparsity prior. Mathematical Problems in Engineering, 2015.

\end{thebibliography}

\end{document}